\documentclass[9pt,twocolumn]{extarticle}
\pdfoutput=1
\usepackage{soul}
\usepackage{titlesec} 

\usepackage[superscript, nomove]{cite}

\usepackage{float}
\usepackage{caption}
\usepackage{lipsum,ulem}
\usepackage[verbose]{placeins}

\usepackage{dsfont}
\usepackage{graphicx}
\usepackage{subcaption}
\usepackage[margin=0.9in]{geometry}
\usepackage[usenames,dvipsnames]{xcolor}
\usepackage[colorlinks,linkcolor=Blue,urlcolor=Blue,citecolor=Blue]{hyperref}
\usepackage{amsmath,amssymb}
\usepackage[squaren]{SIunits}

\usepackage{mathpazo}
\usepackage{courier}
\normalfont
\usepackage[T1]{fontenc}
\usepackage{caption}
\usepackage{multirow}

\usepackage{upgreek}

\newcommand{\ad}[1]{\textsuperscript{#1}\kern-2pt}

\makeatletter
\def\blx@maxline{77}
\makeatother

\usepackage{capt-of}

\def\mytitle{Quantized Landau-level crossing checkerboard in large-angle twisted graphene}

\title{\vspace{-1.0cm}\huge\textbf{\textrm{\mytitle}}}  

\author{Baojuan Dong,$^{1,2,3,4*}$ Kai Zhao,$^{1,2*}$ Kenji Watanabe,$^{5}$ Takashi Taniguchi,$^{6}$ \\ Jianming Lu,$^{7}$ Jianting Zhao$^{8\dagger}$, Fengcheng Wu,$^{9, 10\dagger}$ Jing Zhang,$^{1,2,3\dagger}$ Zheng Han $^{1,2,4\dagger}$}

\date{} 
\begin{document}
\twocolumn[{
\maketitle 
\vspace{-5mm}
\begin{center}
\begin{minipage}{1\textwidth}
\begin{center}
\textit{
\\\textsuperscript{1} State Key Laboratory of Quantum Optics and Quantum Optics Devices, Institute of Optoelectronics, Shanxi University, Taiyuan 030006, P. R. China
\\\textsuperscript{2} Collaborative Innovation Center of Extreme Optics, Shanxi University, Taiyuan 030006, P. R. China
\\\textsuperscript{3} Hefei National Laboratory, Hefei 230088, P. R. China
\\\textsuperscript{4} Liaoning Academy of Materials, Shenyang 110167, P. R. China
\\\textsuperscript{5} Research Center for Electronic and Optical Materials, National Institute for Materials Science, 1-1 Namiki, Tsukuba 305-0044, Japan
\\\textsuperscript{6} Research Center for Materials Nanoarchitectonics, National Institute for Materials Science,  1-1 Namiki, Tsukuba 305-0044, Japan
\\\textsuperscript{7} State Key Laboratory for Mesoscopic Physics, School of Physics, Peking University, Beijing, 100871, P. R. China
\\\textsuperscript{8} Division of Electrical and Magnetic Metrology, National Institute of Metrology, Beijing, P. R. China.
\\\textsuperscript{9} School of Physics and Technology, Wuhan University, Wuhan 430072, P. R. China
\\\textsuperscript{10} Wuhan Institute of Quantum Technology, Wuhan 430206, P. R. China
\vspace{5mm}
\\{$\dagger$} Corresponding to: zhaojt@nim.ac.cn, wufcheng@whu.edu.cn, jzhang74@sxu.edu.cn, vitto.han@gmail.com 
\\{$\star$} These authors contribute equally.
\vspace{5mm}
}
\end{center}
\end{minipage}
\end{center}

\setlength\parindent{13pt}
\begin{quotation}
\noindent 
\section*{Abstract}
{\textbf{When charge transport occurs under conditions like topological protection or ballistic motion, the conductance of low-dimensional systems often exhibits quantized values in units of $e^{2}/h$, where $e$ and $h$ are the elementary charge and Planck's constant. Such quantization has been pivotal in quantum metrology and computing. Here, we demonstrate a novel quantized quantity: the ratio of the displacement field to the magnetic field, $D/B$, in large-twist-angle bilayer graphene. In the high magnetic field limit,  Landau level crossings between the top and bottom layers manifest equal-sized checkerboard patterns throughout the $D/B$-$\nu$ space. It stems from a peculiar electric-field-driven interlayer charge transfer at one elementary charge per flux quantum, leading to quantized intervals of critical displacement fields, (i.e., $\delta D$ = $\frac{e}{2\pi l_{B}^{2}}$, where $l_B$ is the magnetic length). Our findings suggest that interlayer charge transfer in the quantum Hall regime can yield intriguing physical phenomena, which has been overlooked in the past.}}
\end{quotation}
}]

\newpage 
\clearpage

\section*{Introduction}

In a low dimensional quantum conductor, the charge transport is often observed to show discrete characteristics in units of (sometimes with staircases of plateaux under varying magnetic or electrical fields) either superconducting flux quantum defined as $\Phi_{0}=\frac{h}{2e}$, or conductance quantum defined as $G_{0}=\frac{e^{2}}{h}$. However, up to now, only a few candidates in solid states can manifest quantized physical measurables, including the conductance of quantum point contacts \cite{QPC_1,QPC_2,QPC_3,QPC_4}, Shapiro voltage steps in Josephson junctions \cite{ShapiroSteps_1,ShapiroSteps_2}, and the conductance of topologically non-trivial electronic phases such as quantum Hall or Chern insulators \cite{QHE, QSHE_1,QSHE_2, QAHE}. These simple yet elegant kinds of quantization have found key applications in modern quantum technologies like quantum metric standards \cite{QHR_Standards_1,QHR_Standards_2,QHR_Standards_3,QHR_Standards_4}, as well as quantum information processing \cite{QIP}. Exploring previously unexplored systems that may exhibit quantities with quantization (or quantized jumps) is thus of fundamental importance and holds great promise for applications based on exotic quantum states.

Among the reported low dimensional systems, interlayer charge transfer in large angle twisted graphene in the quantum Hall regime has been an interesting system, as the layer-dependent chemical potential is tuned by varying with vertical electric field along a fixed filling fraction $\nu$, charges from the bottom (top) graphene layer migrate to the top (bottom) one \cite{Pablo_PRL,LLs_ZGY,LLs_1,LLs_WL}. Nevertheless, the previous experimental observations are limited to a rather low magnetic field, with little details investigated inside the LL crossing -- possible sophisticated features have been missing since the spin- and valley-degeneracies are not lifted therein \cite{Pablo_PRL,LLs_ZGY,LLs_1,30_philip}. Recently, twisted graphene systems with both small and moderate angles have been investigated \cite{MATBG_SC,MATBG_NEMS,MATBG_CL,MATBG_FCI,MATBG_CI,Ensslin_1,Ensslin_2,LLs_WL,LLs_ZGY}. In those TBLG with twist angles close to the so-called magic angle 1.1 $^{\mathrm{o}}$, flat band physics are seen \cite{MATBG_FCI,MATBG_CI,MATBG_SC,MATBG_NEMS}, while those twisted double bilayer graphene (TDBLG) systems with moderate twist angles of about 2-10 $^{\mathrm{o}}$ show absence of flat bands but the tunable excitonic insulating phases \cite{Ensslin_1,Ensslin_2,LLs_WL,LLs_ZGY}. Electronic transport of TBLG with large twist angles, such as 30$^{\mathrm{o}}$ are also reported in chemical vapor deposited samples \cite{30_philip,PNAS_QC_30,science_QC_30}.

In this work, we show that, by stacking two layers of graphene with a large twist angle (20 - 30 $^{\mathrm{o}}$), specific equal-sized checkerboard patterns can be observed in the quantum Hall regime at the LL crossings throughout the $D$-$n$ space, when all spin- and valley-flavour degeneracies are lifted. Our experimental and theoretical investigations reveal that such quantized 4$\times$4 checkerboard feature originates from the critical electric displacement field $D$, which is bridging the phase transition of two different charge filling states in the TBLG at LL crossings. Such phase transition results from competition between the layer polarization driven by the $D$ field and the capacitance coupling energy, while the charge transfer at LL crossings, driven by vertical electrical fields, corresponds to exactly one elementary charge transferred between the top and bottom Landau orbits. Therefore, at a fixed $\nu$, varying $D$ will lead to LL crossings equally spaced by an interval of $\delta D$ = $\frac{e}{2\pi l_{B}^{2}}=B\cdot e^{2}/h$. We can thus define a measurable, the ratio of $D/B$, which displays quantization of $e^{2}/h$ in the LL crossing checkerboard. Our findings suggest a new paradigm of electric-field-driven interlayer charge transfer in the quantum Hall regime, which produces measurable quantization in $D/B$ and might be a platform for future quantum magnetometry.

\section*{Results}

 \begin{figure*}[t!]
	\centering
	\includegraphics[width=0.99\linewidth]{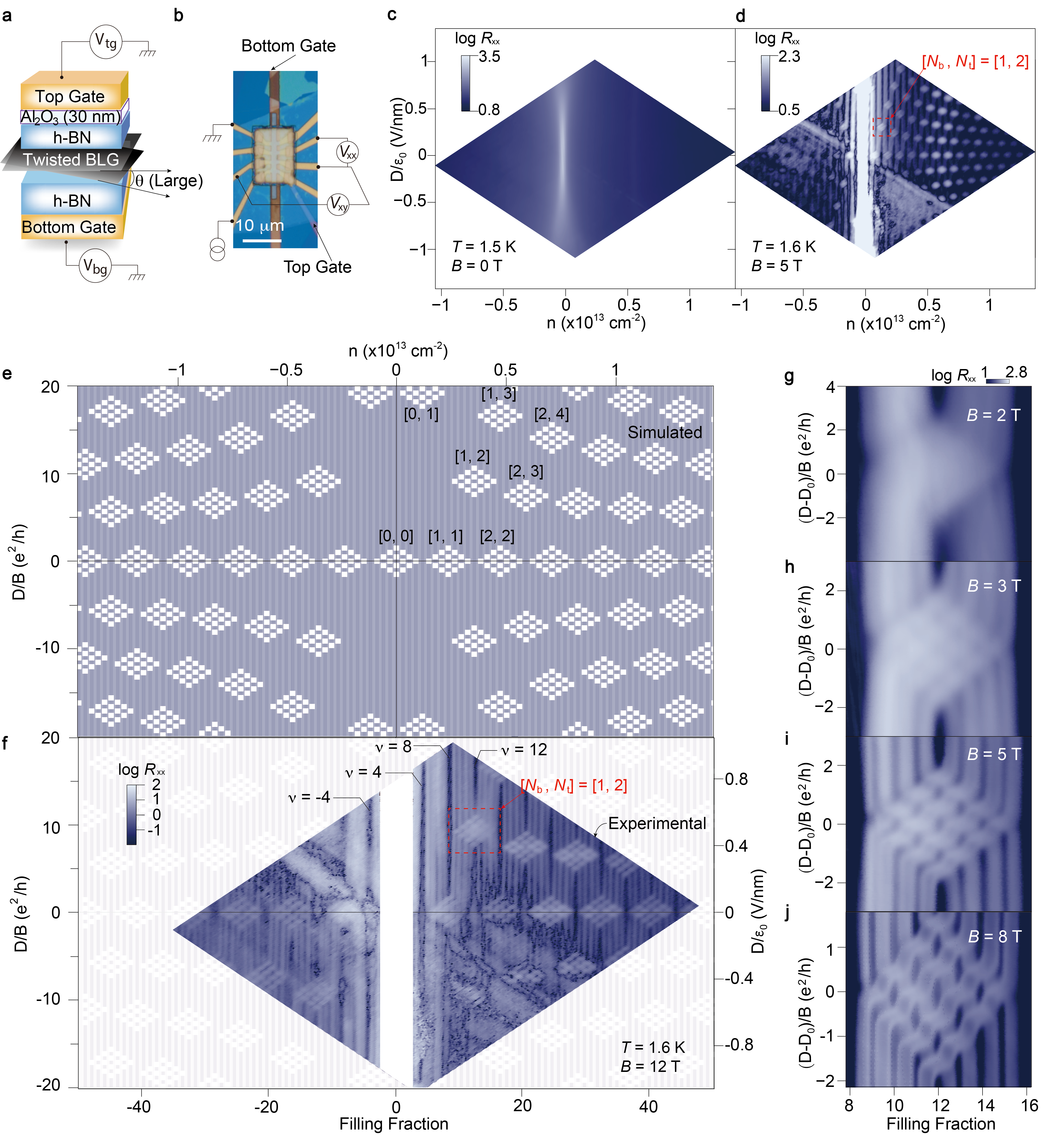}
	\caption{\textbf{Equal-sized of checkerboard cells at LL crossings in the $D$-$n$ space in large angle TBLG.} (a) Schematic illustration of the device. (b) Optical image of a typical LA-TBLG device (Sample-S15, $30^{\mathrm{o}}$-TBLG). Notice that a larger top gate is applied in order to diminish the parasitic contact doping effects. $R_\mathrm{xx}$ mapping in the $D$-$n$ space of Sample-S15 at (c) $T$ = 1.5 K and $B$ = 0 T and (d) $T$ = 1.6 K and $B$ = 5 T (in a log scale for visual clarity). (e) Simulated and (f) experimental checkerboard patterns at LL crossings in the weak layer coupling regime of TBLG at $B$ = 12 T in the $D$-$n$ space. Notice that a background of simulated patterns is overlaid on top of experimental data in (f), for visual guides. Red dashed box indicates LL crossings of [$N_\mathrm{b}$, $N_\mathrm{t}$]=[1, 2]. Typical filling factions of $\nu$ = $\pm$4, 8 and 12 are labelled. $R_\mathrm{xx}$ is truncated at a maximum of 100 $\Omega$, seen as white at the charge neutrality. (g)-(j) show the evolution from the single-dotted to 4$\times$4 checkerboard-like LL crossings at [$N_\mathrm{b}$, $N_\mathrm{t}$]=[1, 2], at $B$ = 2, 3, 5 and 8 T, respectively. Data were obtained at $T$ = 1.6 K.}
\end{figure*}

\begin{figure*}[t!]
	\centering
	\includegraphics[width=0.95\linewidth]{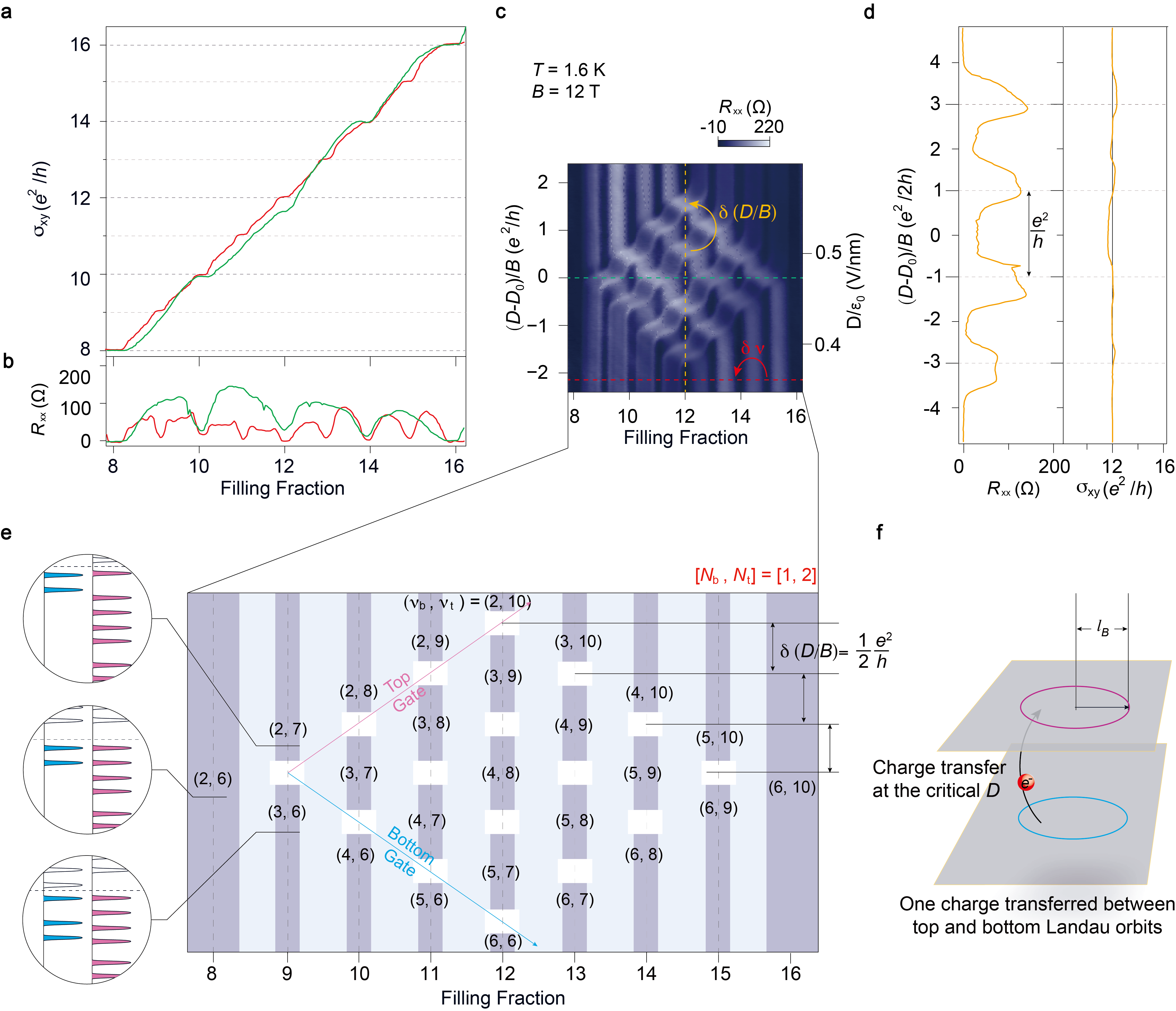}
	\caption{\textbf{Quantized $D/B$ jumps of interlayer charge transfer phase transition at fixed $\nu$ in the LL crossing area of $[N_{b}, N_{t}]=[1, 2]$.} (a) and (b) plot the line profiles of $\sigma_\mathrm{xy}$ and $R_\mathrm{xx}$ along green and red dashed lines in (c). The same colour codes are used for the solid lines in (a)-(b), and the dashed lines in (c). The $y$-axis of colour mapping of $R_\mathrm{xx}$ in (c) is plotted in both raw $D$ (right, corrected according to Supplementary Figure 10 and Supplementary Note 1) and $(D-D_{0})/B$ (left), where $D_{0}$ is defined as the central $D$ of the checkerboard. It is noticed that, when plotted in $(D-D_{0})/B$, adjacent LL crossings are found to be separated by a quantized value, in a unit of $e^{2}/h$ when $\nu$ is fixed. (d) illustrates line profiles along $\nu$ = 12, i.e., the orange dashed line in (c), of $R_\mathrm{xx}$ and $\sigma_\mathrm{xy}$, respectively. (e) Art drawing of the quantized 4$\times$4 checkerboard at the LL crossing of $[N_{b}, N_{t}]=[1, 2]$. The fillings of Landau bands for (2, 6), (3, 6), and (2, 7) are schematically illustrated on the left side of the panel. (f) A cartoon illustration for the event of charge transfer at critical $D$ for any phase boundaries (marked as white boxes) along fixed $\nu$ in (e).}
\end{figure*}

\begin{figure*}[ht!]
	\centering
	\includegraphics[width=0.95\linewidth]{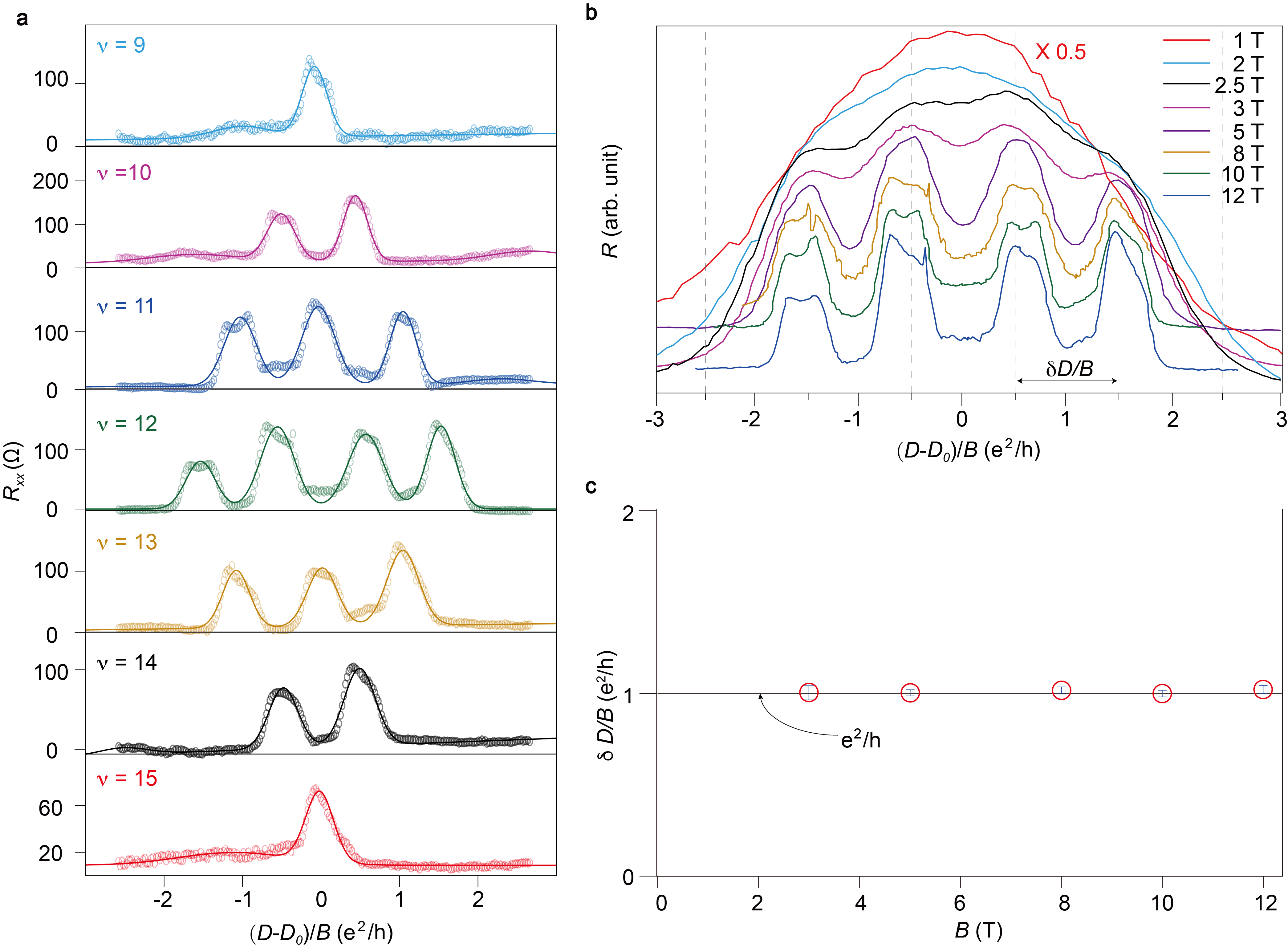}
	\caption{\textbf{Magnetic field dependence of $D/B$ quantizations at fixed $\nu$ in the LL crossing area of $[N_{b}, N_{t}]=[1, 2]$.} (a) The line profiles of $R_\mathrm{xx}$ at 1.6 K with 12 T magnetic field at filling factors $\nu$ from 9 to 15, which were fitted by using Gaussian method to extract the precise crossing $R_\mathrm{xx}$ peak positions in the checkerboard. (b) $R_\mathrm{xx}$ at $\nu = 12$  as a function of $(D-D_{0})/B$ at different magnetic fields from 1 to 12 T. Data are shifted in the $y$-axis for visual clarity. (c) The statistics of $\delta (D/B)$ as a function of magnetic field. Quantization at $e^{2}/h$ of $\delta (D/B)$ is clearly seen in (b). Error bars in (c) are defined as the standard deviation of nine values of $\delta (D/B)$  at each magnetic field.}.
	\label{fig:fig3}
\end{figure*}

\noindent\textbf{Fabrications and characterizations of large angle TBLG devices.} \\ Monolayered graphene and few-layered hexagonal boron nitride (h-BN) flakes were exfoliated from bulk crystals. The graphene was cut by a conductive atomic force microscope (AFM) tip, using the anode-oxidation technique \cite{AFM,AFM_1}. The AFM-cut graphene twin flakes were then stacked with a twist angle of 20$^\mathrm{o}$ or 30$^\mathrm{o}$ using the dry transfer method \cite{Lei_Science_2013}, and then encapsulated by top and bottom h-BN flakes. Detailed fabrication processes can be seen in Supplementary Figures 1-3. The device were equipped with dual gates and electrodes of Ti/Au via standard lithography and electron-beam evaporation (fabrication details are available in Methods), as illustrated in Fig. 1a. Fig. 1b shows the optical micrograph of a typical large-angle (LA) TBLG device (Sample-S15, 30 $^\mathrm{o}$-twisted), with the corresponding fabrication flow shown in Supplementary Figure 2. Before further analysis, we define the two gates induced displacement field $D = (C_\mathrm{tg}V_\mathrm{tg}-C_\mathrm{bg}V_\mathrm{bg})/2 - D_{r}$, and the total carrier density $n_\mathrm{tot}=(C_\mathrm{tg}V_\mathrm{tg}+C_\mathrm{bg}V_\mathrm{bg})/e-n_{r}$, as commonly used in dual-gated graphene devices \cite{FengWang_Nature_BLG, Maher_Science}. Here, $C_\mathrm{tg}$ and $C_\mathrm{bg}$ are the top and bottom gate capacitances per area, respectively. And $V_\mathrm{tg}$ and $V_\mathrm{bg}$ are the top and bottom gate voltages, respectively. $n_{r}$ and $D_{r}$ are residual doping and residual displacement field, respectively. Fig. 1c shows a mapping of longitudinal resistance $R_\mathrm{xx}$ of Sample-S15 in the $D$-$n$ space, when cooled down to $T$ = 1.5 K at $B$ = 0 T. It is noteworthy that, along the charge neutral line (carrier density $n$ = 0), $R_\mathrm{xx}$ decreases when $D$ is departing from zero. This behaviour speaks of the weak layer coupling nature \cite{Pablo_PRL}, which largely distinguishes from conventional strongly-coupled Bernal-stacked bilayer graphene system. In the latter, $R_\mathrm{xx}$ at the charge neutrality monotonously increases upon increasing the absolute value of $D$, due to the gap opening at the Dirac point \cite{bilayer_KNY,bilayer_Pablo}. Data from control sample of Bernal-stacked bilayer graphene are compared in a side-by-side manner in Supplementary Figure 4.

Fig. 1d illustrates the mapping of $R_\mathrm{xx}$ (in a log scale for visual clarity) in the $D$-$n$ space within the same range of Figure 1c at $B$ = 5 T. Clearly, resistive states (circular-shaped dots) can be seen at each LL crossing in both electron and hole sides, while each dot is non-identical with a dispersion in their sizes. This is in echo with the previous observation in a weakly-coupled (with unreported twist-angle) TBLG sample \cite{Pablo_PRL}. Notice that here all the LL crossing are with 4-fold degeneracy, as can be seen in the transverse conductance $\sigma_\mathrm{xy}$ in Supplementary Figure 5, when plotted in the $D$-$n$ space. Hence, all the crossing points can be marked by the Landau index $[N_\mathrm{b}, N_\mathrm{t}]$, as shown in Supplementary Figure 5b.

When further increasing the perpendicular magnetic field, spin- and valley-degeneracy start to be lifted, and the single-dot-shaped resistive states in the $D$-$n$ space at low $B$ will develop into 4$\times$4 matrices. The newly developed LL crossings can be labelled by ($\nu_\mathrm{b}$, $\nu_\mathrm{t}$), within a given [$N_\mathrm{b}$, $N_\mathrm{t}$], where $\nu_\mathrm{b,t}$ is the filling fraction of each layer. Surprisingly, it is observed that all 4$\times$4 checkerboard cells exhibit the same uniform size throughout the $D$-$n$ space, regardless of their parent index [$N_\mathrm{b}$, $N_\mathrm{t}$], which is the central finding of this work, as shown in Fig. 1e-f. Here, Fig. 1e is the theoretical simulation that agrees well with the experimental data in Fig. 1f. And the $x$- and $y$-axis of Fig. 1e-f are linearly transformed from $n$ to $\nu$, and from $D$ to $D/B$, respectively. More theoretical details will be discussed in  the coming sections. In another typical LA-TBLG device (Sample-S37, 20 $^\mathrm{o}$-twisted), we carried out high-resolution mapping of $R_\mathrm{xx}$ in the $D$-$\nu$ space, at [$N_\mathrm{b}$, $N_\mathrm{t}$]=[1, 2] (indicated by dashed red box in Fig. 1f), for different magnetic fields. As shown in Fig. 1g-j, the evolution from the single-dotted to 4$\times$4 checkerboard-like LL crossings can be clearly seen, from 2 T to 8 T, which is consistent with the observations in Sample-S15, as shown in Supplementary Figure 6. We show in  Supplementary Figures 7-9 other tested LA-TBLG samples.

\begin{figure*}[ht!]
	\centering
	\includegraphics[width=0.95\linewidth]{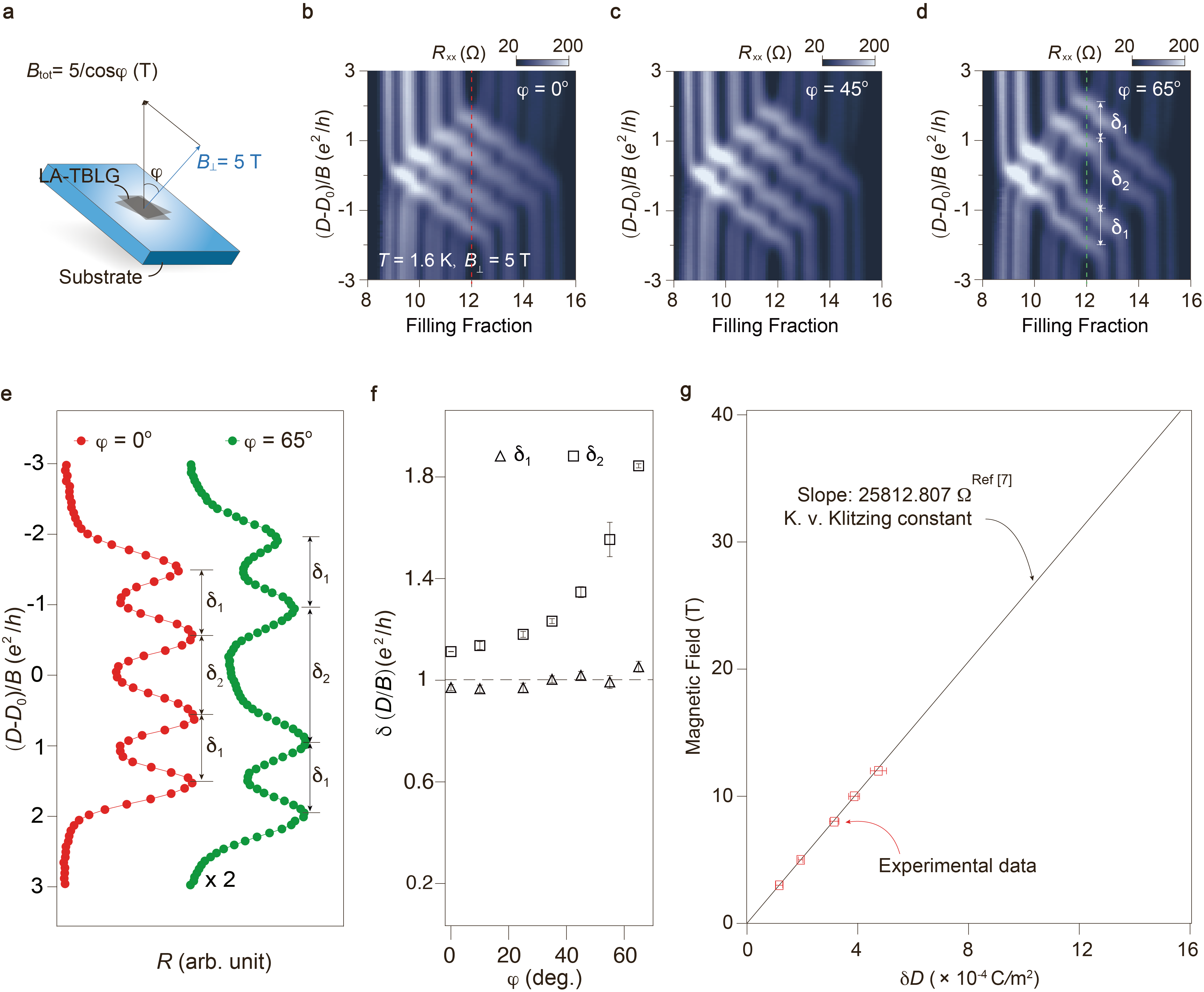}
	\caption{\textbf{Tuning the LL-crossing checkerboards with Zeeman energy.} (a) A schematic drawing of a LA-TBLG sample measured under tilted magnetic field. A total magnetic field is varied to maintain the vertical filed $B_\mathrm{\perp}$ at 5 T. (b)-(d) are colour mapping of $R_\mathrm{xx}$ plotted in the parameter space of $(D-D_{0})/B$ and filling fractions, with the tilt angle $\varphi$ being 0$^\mathrm{o}$, 45$^\mathrm{o}$, and 65$^\mathrm{o}$, respectively. All data are obtained at $T$= 1.6 K, and $B_{\perp}$ = 5 T. (e) Line profiles of $R_\mathrm{xx}$ at $\nu = 12$ in (b) (red) and (d) (black), respectively. The two curves are shifted in the $x$-axis for visual clarity. (f) Two typical sets of $\delta D$ (the difference between displacement fields for adjacent resistive peaks in a LL-crossing checkerboard), i.e., $\delta_{1}$ and $\delta_{1}$ as indicated in (d) and (e), are plotted as a function of tilt angle. (g) The magnetic field $B$ plots against $\delta D$ in the unit of C/m$^{2}$. The slope is then equal to the K. v. Klitzing constant. This linear dependence of $B$-$\delta D$, once calibrated using experimental data at relatively low $B$ can then be used to measure unknown $B$ at high $B$ and low $T$, by reading the $\delta D$ in the quantized LL-crossing checkerboard.}.
	\label{fig:fig4}
\end{figure*}

\vspace{5mm}
\noindent\textbf{Quantized $D/B$ jumps.} In this study we mainly focus on the physical origin of the equally sized checkerboard patterns at the LL crossing, taking the [$N_\mathrm{b}$, $N_\mathrm{t}$]=[1, 2] checkerboard as an example for examination. The displacement field used in all the figures and discussions has been corrected by using the decoupled interlayer model, taking into account the sample-dependent interlayer quantum capacitance $C_\mathrm{GG}$ (which largely influences the calculated $D$) of the TBLG \cite{Pablo_PRL}. In Supplementary Note 1 and Supplementary Figure 10, we show that using the decoupled TBLG model \cite{Pablo_PRL}, $C_\mathrm{GG}$ can be fitted to be 6.30 $\mu$F/cm$^{2}$ for Sample-S37. As a comparison, $C_\mathrm{GG}$ of Sample-S15 is also calibrated in Supplementary Figure 10. Figure 2a and 2b plot the line profiles of $\sigma_\mathrm{xy}$ and $R_\mathrm{xx}$ along green and red dashed lines in Fig. 2c, while the same colour codes are used for the line profiles.  At $D/\varepsilon_{0}$ = 0.3493 V/nm (outside the checkerboard), spin and valley degeneracies are fully lifted, with all integer quantum Hall states from filling  8 to 16 observed as quantized conductance plateau in $\sigma_\mathrm{xy}$, and corresponding resistance minima in $R_\mathrm{xx}$. On the other hand, at $D/\varepsilon_{0}$ = 0.4726 V/nm (centre of the checkerboard), only even-integer quantum Hall states are seen in $\sigma_\mathrm{xy}$. The odd integer fillings are occupied by the resistive states at LL crossings.

Interestingly, when scaling the displacement field into $D/B$, all the resistive states at LL crossings are found to be located at a quantized value in the $(D-D_{0})/B$ axis with respect to the centre of the checkerboard, in units of $e^{2}/2h$. This effect can be more pronounced when plotted as line profiles along $\nu$ = 12 of both $R_\mathrm{xx}$ and $\sigma_\mathrm{xy}$, with the $y$-axes shifted to the centre of the checkerboard, noted as $(D-D_{0})/B$, shown in Fig. 2d. We also rescaled all the $y$-axes into $D/B$ for Fig. 1e-f, or $(D-D_{0})/B$ for Fig. 1g-j, respectively.

In the LL-crossing checkerboard plotted in the parameter space of $(D-D_{0})/B$ and filling fraction $\nu$, two types of quantized properties can be observed, with one of them being the well known QHE-originated quantization of filling fractions, and the other being the interval in $(D-D_{0})/B$ for $R_\mathrm{xx}$ peaks, as indicated by the arrows in Fig. 2c. 
We now provide a theoretical explanation for this observation.
For low-energy electrons, our current system can be approximated by two layers that are decoupled at the single particle level because of the large twist angle, but capacitively coupled by Coulomb interaction. Due to the spin and valley degeneracies within each graphene layer, the filling factor $\nu_{i}$ and Landau level index $N_{i}$ have the following relation, $\nu_i=4N_i-2+\widetilde{\nu}_i$, where $i$ is the layer index, $\widetilde{\nu}_i$ is the filling factor of the partially-filled Landau level with index $N_{i}$, and $0\leq \widetilde{\nu}_i\leq 4$. Hence, the filling factor within the LL-crossing checkerboard indexed by [$N_\mathrm{b}, N_\mathrm{t}$] can be an integer number between $4(N_\mathrm{b}+N_\mathrm{t})\pm 4$, also written as $\nu=\nu_\mathrm{b}+\nu_\mathrm{t}=4(N_\mathrm{b}+N_\mathrm{t}-1)+\widetilde{\nu}$, where $\widetilde{\nu}=\widetilde{\nu}_\mathrm{b}+\widetilde{\nu}_\mathrm{t}$.   We construct the total energy per area (defined as $E_\mathrm{tot}$) that comprises three terms, the single-particle energy of the occupied LLs, the classical electrostatic energy (including the layer potential difference generated by the $D$ field and the capacitance energy), and the intralayer exchange energy, as described in Methods. Theoretically, we assume that the integer quantum Hall insulators are formed within each layer, neglecting the interlayer exchange interaction and Zeeman energy (the subtle effects of spin ordering and Zeeman effect are discussed in the Methods). 

In the LL-crossing checkerboard indexed by [$N_\mathrm{b}, N_\mathrm{t}$], the states at a fixed $\nu$ can change from one phase $\{N_\mathrm{b}, \widetilde{\nu}_\mathrm{b}; N_\mathrm{t}, \widetilde{\nu}_\mathrm{t}\}$ to another adjacent one $\{N_\mathrm{b},\widetilde{\nu}_\mathrm{b}-1;N_\mathrm{t},\widetilde{\nu}_\mathrm{t}+1\}$ driven by the displacement field, leading to a phase transition that yields discrete phase boundaries (corresponding to experimentally observed resistive peaks at a given $\nu$ in the checkerboards). The critical displacement field can then be found at the conditions when the energies of the above two adjacent filling phases become degenerate, as shown in Methods. Solving them, one can obtain all the critical field $\widetilde{D}=\frac{D}{e n_0}=\frac{D}{B}\Big/\Big(\frac{e^2}{h}\Big)$ in each checkerboard to be 0, $\pm 1/2$, $\pm 1$ and $\pm 3/2$ measured relative to the centre of the checkerboard, as given in the third column in Table 1 in the Methods. At this stage, it is clear that $\delta D/B$ between each adjacent phase boundary at fixed $\nu$ is quantized to $e^{2}/h$.
As our theory shows, the value of $\delta D/B$ within a checkerboard is fully determined by the competition between the $ D$-field-driven layer potential difference and the Coulomb-driven capacitance energy, where the former (latter) favors layer polarization (equal charge distribution between the two layers), which makes $\delta D/B$ independent of the parent index [$N_\mathrm{b}, N_\mathrm{t}$]. As integer quantum Hall insulators are incompressible with an integer number of electrons per flux quantum, the charge transfer at the crossings is exactly one charge per Landau orbit of area $2\pi l_B^2$, making $\delta D$ quantized to $\frac{e}{2\pi l_B^2}$.
Using the experimentally extracted $C_\mathrm{GG}$, we plot the full map of checkerboards in the space of $D/B$ and $\nu$ in Fig. 1e (and the background in Fig. 1f), which agrees well with experimental results. 

Figure 2e illustrates the theoretical picture discussed above, taking the LL-crossing indexed by [$N_\mathrm{b}, N_\mathrm{t}$]=[1, 2] as an example. White boxes denote the phase boundaries at which the states change along the vertical direction (i.e., the fixed $\nu$), driven by the displacement field. For the three filling states ($\nu_\mathrm{b}, \nu_\mathrm{t}$) = (2, 6), (3, 6) and (2, 7), LLs alignments for each layer are illustrated in the circled schematics (blue and pink colour denote LLs for bottom and top layers, respectively). Those white boxes form an evenly distributed matrix with 4$\times$4 elements, due to the quantization of both $\nu$ in the $x$-axis, and $D/B$ in the $y$-axis, respectively. The latter quantization of the measurable $D/B$ within a checkerboard at LL-crossing was not quantitatively established before, although similar matrices of resistive peaks in LL-crossing of twisted graphene have been reported elsewhere \cite{LLs_WL}. Charge transfer between LLs in two sub-bands system of double quantum wells was also investigated, however, quantization signature reported here was neither seen therein \cite{LiuyangPRL2011}. In experiment, the two solid blue and pink arrows in Fig. 2e indicates the sweeping direction in terms of bottom and top gates, respectively (see also Supplementary Figure 6d). The quantization of $\delta D/B$ is thus a manifestation of the well-defined interlayer distance in LA-TBLG with well-developed quantum Hall ferromagnetism of spin and valley ordering, studied in the current work.

We further investigate the $B$ and $T$-dependencies of $D/B$ quantizations at fixed $\nu$ within one quantized checkerboard. Fig. 3a depicts the line profiles of $R_\mathrm{xx}$ for $\nu$ from 9 to 15, for the checkerboard at LL-crossing of [$N_\mathrm{b}, N_\mathrm{t}$]=[1, 2] measured in Sample S37 at $T$ = 1.6 K and $B$ = 12 T (same analysis of Sample S15 can be found in Supplementary Figure 11). These resistive peaks at critical $D$ can be fitted using the Gaussian distribution, and hence the distance between each pair of adjacent peaks can be determined as $\delta D/B$. We plot the $R_\mathrm{xx}$ at $\nu$ = 12 as a function of magnetic fields, as shown in Fig. 3b. It is seen that, at relatively low $B$, the LL-crossing are still in the stage of a single dotted resistive peak, as no degeneracy lifting has been possible. Above about 3 T, clear 4-peak feature can be observed in the line profile, and peaks are well developed at about 5 T.  Fig. 3c illustrates the measured $\delta D/B$ (in units of $e^{2}/h$) as a function of magnetic field, and the experimental data distribute around the unity, which is expected according to our theoretical modelings. The data in Fig. 3c is derived from the statistical analysis of nine $\delta D/B$ values, as detailed in Supplementary Tables 1 and 2. More data on the temperature dependence of $(D-D_{0})/B$ are given in Supplementary Figure 12.

\bigskip
\noindent\textbf{Tuning the LL-crossing checkerboard with Zeeman energy.} At this stage, it is noticed that the uniform 4$\times$4 matrix checkerboard patterns were observed in only a small fraction of (fewer than 1/10) the as-fabricated samples, including such as Sample-S15 and S37. As shown in the Supplementary Figures 9-11, the checkerboard patterns in other samples were distorted, where these resistive states (crossing points in the 4$\times$4 matrix) are divided into four sub-groups. Our theory described in the Methods, which assumes spin conservation in the charge transfer, is found to be well-matching the data of uniform 4$\times$4 matrix checkerboards. However, different samples may exhibit different spin orderings due to the competition between Zeeman energy and the atomic scale interactions. To examine the Zeeman effect, we performed tilted-field magnetotransport of the Sample S37. As illustrated in Fig. 4a, the LA-TBLG sample is subjected to a tilted field, while the total magnetic field ($B_\mathrm{tot}$) is increased with $B_\mathrm{\perp}$ held constant. The in-plane projection $B_\mathrm{\parallel}$ = $B_\mathrm{tot} \sin \varphi$ can thus effective tune the Zeeman energy in the tested system, where $\varphi$ is the angle of tilt. 

As shown in Fig. 4b-d, at a fixed index of [$N_\mathrm{b}, N_\mathrm{t}$]=[1, 2] and a constant $B_\mathrm{\perp}$ = 5 T at $T$ = 1.6 K, the 4$\times$4 matrix checkerboards is tuned from its initial uniform distribution (Fig. 4b, $\varphi$ = 0$^\mathrm{o}$) into gradually distorted patterns (Fig. 4c, $\varphi$ = 45$^\mathrm{o}$). For the case of $\varphi$ = 65$^\mathrm{o}$ in Fig. 4d, the checkerboard is clearly divided into four sub-groups, with the differences between adjacent $R_\mathrm{xx}$ in the axis of $(D-D_{0})/B$, defined as $\delta D/B$, developing into two typical values, marked as $\delta_{1}$ and $\delta_{2}$, respectively, as indicated in Fig. 4d. Line profiles of $R_\mathrm{xx}$ against $(D-D_{0})/B$ at $\nu = 12$ for the case of tilted field angle $\varphi$ = 0$^\mathrm{o}$ (red dashed line in Fig. 4b) and $\varphi$ = 65$^\mathrm{o}$ (green dashed line in Fig. 4d) are plotted in Fig. 4e, with the $R_\mathrm{xx}$ value shifted for visual clarity. With the in-plane magnetic field $B_\mathrm{\parallel}$ varied from zero (red) to 10.72 T ($\varphi$ = 65$^\mathrm{o}$, green), the value of $\delta_{2}$ is almost doubled. Figure 4f further plots the $\delta D/B$ as a function of $\varphi$. While $\delta_{2}$ increases drastically with the tilt angle, $\delta_{1}$  remains remarkably quantized at $e^2/h$. This is direct evidence that the checkerboard pattern can be effectively influenced by tuning the Zeeman energy in the system. Despite this influence, the quantization of $\delta D/B$ in $e^2/h$ can still be manifested, for example, through $\delta_{1}$.   Further detailed theory that takes into account different spin orders will be needed to quantitatively describe such subtle behaviors.

Finally, we show a potential application using the phenomenon of uniform 4$\times$4 matrices of quantized LL-crossing checkerboards for LA-TBLG devices. When plotting the magnetic field $B$ with a unit of $\textrm{T}$ against $\delta D$ with a unit of $\textrm{C}/\textrm{m}^2$, it should yield a slope equal to the K. v. Klitzing constant $h/e^{2}$, as shown in Fig. 4g. Intriguingly, such a linear dependence of $B$-$\delta D$, once calibrated using experimental data at relatively low $B$, can then be retained in the device, which can be further used as a kind of quantum magnetometer for measuring unknown $B$ at high $B$ and low $T$, by simply reading the $\delta D$ in the quantized LL-crossing checkerboard.

\bigskip

To conclude, we have devised a system of large-angle ($20^{\mathrm{o}}$ -  $30^{\mathrm{o}}$) twisted bilayer graphene, in which  equal-sized 4$\times$4 checkerboards are seen at each interlayer LL-crossing point throughout the $D$-$n$ parameter space, when all spin- and valley-flavour degeneracies are lifted. When looking into one of such checkerboards in the $D$-$n$ (or $D/B$ - $\nu$) space, in addition to the well known QHE-driven quantized properties in the $\nu$ axis, varying $D$ at a fixed integer filling fraction $\nu$ will also yield quantized distance between resistance peaks in the $D/B$ axis in a unit of $e^{2}/h$. Such quantization of $\delta D/B$ originates from the charge quantization per Landau orbit of integer quantum Hall states. Our findings of the long-overlooked quantized property in the measurable $D/B$ suggest that electric-field-driven interlayer charge transfer in the quantum Hall regime in a double-layer 2D electronic system may be intriguing when realized in condensed matter devices.


\clearpage

 \clearpage

\section*{Methods}
\vspace{3mm}
\noindent\textbf{Sample fabrication.} vdW few-layers of the h-BN/graphene/h-BN sandwich were obtained by mechanically exfoliating high quality bulk crystals. The vertical assembly of vdW layered compounds were fabricated using the dry-transfer method in an ambient conditions. The h-BN/graphene/h-BN sandwich were then transferred onto the pre-fabricated h-BN/Au substrate. Hall bars of the devices were achieved by reactive ion etching. During the fabrication processes, electron beam lithography was done using a Zeiss Sigma 300 SEM with a Raith Elphy Quantum graphic writer. One-dimensional edge contacts were achieved by using the electron beam evaporation. After atomic layer deposition of about 30 nm Al$_{2}$O$_{3}$, big top gate was deposited to form the complete dual gated h-BN encapsulated large angle twisted bilayer graphene devices as shown in Fig. 1a and b. Gate electrodes as well as contacting electrodes were fabricated with a electron beam evaporation, with typical thicknesses of Au/Ti $\sim$ 30/5 nm and $\sim$ 50/5 nm, respectively.

\vspace{3mm}
\noindent\textbf{AC Electrical measurements.} During measurements, the graphene layers were fed with an AC $I_\mathrm{bias}$ of about 50 - 500 nA. The longitudinal and Hall voltages were recorded using low-frequency SR830 lock-in amplifiers. Four-probe measurements were used throughout the transport measurements under high magnetic field and at low temperatures in an Oxford TeslaTron cryostat. Gate voltages on the as-prepared Hall bar devices were maintained by a Keithley 2400 source meter.

\vspace{3mm}
\noindent\textbf{Theory of Landau-Level Crossing Checkerboards}.

\textbf{I. Energetics.} We formulate the energy of the LA-TBLG in a strong magnetic field. Because of the large twist angle, the low-energy electronic states of the two layers can be described by decoupled Dirac fermions at the single-particle level. In the presence of the magnetic field, the Dirac fermions form quantized Landau levels. We denote the Landau level filling factor of the $i$ layer as $\nu_{i}$, where $i=b$ and $t$, respectively, for the bottom ($b$) and top ($t$) layer. We use $N_{i}$ to label the index of the partially-filled Landau level nearby the Fermi energy in the $i$ layer. Given the spin and valley degrees of freedom within each graphene layer, $\nu_{i}$ and  $N_{i}$ have the following relation,
\begin{equation}
\nu_i=4N_i-2+\widetilde{\nu}_i,
\end{equation}
where $\widetilde{\nu}_i$ is the filling factor of the partially-filled Landau level with index $N_{i}$, and $0\leq \widetilde{\nu}_i\leq 4$. 

The electron density of the $i$ layer is $n_i=\nu_i n_0$. Here $n_0$ is the electron density per Landau level given by 
\begin{equation}
    n_0=\frac{1}{2\pi l^2_B}=\frac{eB}{h},
\end{equation}
where $l_B=\sqrt{\hbar/(eB)}$ is the magnetic length.

The total filling factor is
\begin{equation}
\nu_{\text{tot}}=\nu_\mathrm{b}+\nu_\mathrm{t}=4(N_\mathrm{b}+N_\mathrm{t}-1)+\widetilde{\nu},
\end{equation}
where $\widetilde{\nu}=\widetilde{\nu}_\mathrm{b}+\widetilde{\nu}_\mathrm{t}$.

In our theory, the system can be described using four indices $\left\{ N_\mathrm{b},\widetilde{\nu}_\mathrm{b};N_\mathrm{t},\widetilde{\nu}_\mathrm{t} \right\}$ in the quantum Hall regime, which determines the filling factor of each layer and the total filling factor. 

The total energy per area $E_{\text{tot}}$ of the system in the quantum Hall regime can be decomposed into the following contributions,
\begin{equation}
E_{\text{tot}}[N_\mathrm{b},\widetilde{\nu}_\mathrm{b};N_\mathrm{t},\widetilde{\nu}_\mathrm{t}]=E_{\text{LL}}+E_{\text{C}}+E_{\text{X}}.
\label{Etot}
\end{equation}

First, $E_{\text{LL}}$ accounts for the single-particle energies of the occupied Landau levels, 
\begin{equation}
E_{\text{LL}}=\sum_{i=b,t}\left\{
\sum_{N=-\infty}^{N_i-1}
4\mathcal{E}_{\text{LL}}(N)n_0+
\mathcal{E}_{\text{LL}}(N_i)\widetilde{\nu}_i n_0
\right\}, 
\end{equation}
where $\mathcal{E}_{\text{LL}}(N)=\text{sgn}(N)v_F \sqrt{2\hbar eB |N|}$ is the energy of the $N$th Landau level, with $v_F$ being the velocity of the Dirac fermion. 

Second, $E_{\text{C}}$ describes the classical electrostatic energy for the charge distribution of the bilayer
\begin{equation}
E_{\text{C}}=\frac{U}{2}(n_\mathrm{b}-n_\mathrm{t})-\frac{e^2d_{\text{GG}}}{2\varepsilon_{\text{GG}}}n_\mathrm{t} n_\mathrm{b}
=E_0 [\widetilde{D}(\nu_\mathrm{b}-\nu_\mathrm{t})-\nu_\mathrm{t} \nu_\mathrm{b}],
\label{EC}
\end{equation}
where $U=eDd_{\text{GG}}/\varepsilon_{\text{GG}}$ is the layer potential difference generated by the displacement field $D$, $e>0$ is the elementary charge, $d_{\text{GG}}$ is the interlayer distance, and $\varepsilon_{\text{GG}}$ is the dielectric constant. The second term  in $E_{\text{C}}$ arises from the capacitor energy of the charged bilayer, where the capacitance per area is $C_{\text{GG}}=\varepsilon_{\text{GG}}/d_{\text{GG}}$. In Eq.~\eqref{EC}, $E_0=e^2n_0^2/(2C_{\text{GG}})$ sets a scale for the electrostatic energy, and $\widetilde{D}$ is given by,
\begin{equation}
\widetilde{D}=\frac{D}{e n_0}=\frac{D}{B}\Big/\Big(\frac{e^2}{h}\Big),
\label{tildeD}
\end{equation}
where $\widetilde{D}$ is dimensionless. 

Finally, $E_{\text{X}}$ is the intralayer exchange energy of the filled Landau levels. We assume that $E_{\text{X}}$ takes the following form,
\begin{equation}
E_{\text{X}}=\sum_{i=b,t}\left\{(4-\widetilde{\nu}_i)\mathcal{E}_{\text{X}}(N_i-1)+ \widetilde{\nu}_{i} \mathcal{E}_{\text{X}}(N_i) \right\}.    
\end{equation}
The assumption is as follows. Because $\widetilde{\nu}_{i}$ flavours occupy Landau levels with index $N\le N_{i}$, they contribute   exchange energy of $\widetilde{\nu}_{i} \mathcal{E}_{\text{X}}(N_i)$ with $\mathcal{E}_{\text{X}}(N_i)$ being the exchange energy per flavour. A similar reasoning leads to the other term $(4-\widetilde{\nu}_i)\mathcal{E}_{\text{X}}(N_i-1)$. We assume that the exchange energy is additive regarding the flavour degree of freedom, which is reasonable because of the approximate SU(4) symmetry of graphene in the quantum Hall regime.

Motivated by the experimental observation, we focus on the case that both $\nu_\mathrm{b}$ and $\nu_\mathrm{t}$ are integers. We do not consider the case of the interlayer coherent state. Therefore, integer quantum Hall insulators are formed within each layer, and interlayer exchange interaction can be neglected. 
On the other hand, the Hartree energy is already included in the electrostatic energy of $E_{\text{C}}$.

We note that the Zeeman energy is not included in Eq.~\eqref{Etot}. We discuss the effect of the spin ordering and Zeeman energy separately. 

\textbf{II. Critical displacement fields}. At a critical displacement field for the Landau level crossing, the energies of two different states become degenerate. When the crossing checkerboard is formed, the filling factor within each layer changes by 1 at each critical displacement field.  Therefore, we compare the energies of two states described respectively by $\{N_\mathrm{b},\widetilde{\nu}_\mathrm{b}-1;N_\mathrm{t},\widetilde{\nu}_\mathrm{t}+1\}$  and  $\{N_\mathrm{b},\widetilde{\nu}_\mathrm{b};N_\mathrm{t},\widetilde{\nu}_\mathrm{t}\}$,
\begin{equation}
\begin{split}
&E_{\text{tot}}[N_\mathrm{b},\widetilde{\nu}_\mathrm{b}-1;N_\mathrm{t},\widetilde{\nu}_\mathrm{t}+1]
-E_{\text{tot}}[N_\mathrm{b},\widetilde{\nu}_\mathrm{b};N_\mathrm{t},\widetilde{\nu}_\mathrm{t}]\\
=&[\mathcal{E}_{\text{LL}}(N_\mathrm{t})
-\mathcal{E}_{\text{LL}}(N_\mathrm{b})]n_0
+E_0[-2\widetilde{D}+1+4(N_\mathrm{t}-N_\mathrm{b})+\widetilde{\nu}_\mathrm{t}-\widetilde{\nu}_\mathrm{b}]\\
&+[ \mathcal{E}_{\text{X}}(N_\mathrm{t})-\mathcal{E}_{\text{X}}(N_\mathrm{t}-1))]-[ \mathcal{E}_{\text{X}}(N_\mathrm{b})-\mathcal{E}_{\text{X}}(N_\mathrm{b}-1))].
\end{split}
\label{energycompare}
\end{equation}

By setting Eq.~\eqref{energycompare} to zero, we obtain the critical displacement field, which is  given by,
\begin{equation}
    \widetilde{D}_{N_\mathrm{b},N_\mathrm{t}}^{(c)}(\widetilde{\nu}_\mathrm{b},\widetilde{\nu}_\mathrm{t} \leftrightarrow \widetilde{\nu}_\mathrm{b}-1,\widetilde{\nu}_\mathrm{t}+1)=\widetilde{D}_{N_\mathrm{b},N_\mathrm{t}}^{(0)}+\frac{\widetilde{\nu}_\mathrm{t}-\widetilde{\nu}_\mathrm{b}+1}{2},
\label{criticalD}
\end{equation}
where $\widetilde{D}_{N_\mathrm{b},N_\mathrm{t}}^{(0)}$ is independent of $\widetilde{\nu}_\mathrm{b}$ and $\widetilde{\nu}_\mathrm{t}$,
\begin{equation}
\begin{split}
&\widetilde{D}_{N_\mathrm{b},N_\mathrm{t}}^{(0)}=\{[\mathcal{E}_{\text{LL}}(N_\mathrm{t})-\mathcal{E}_{\text{LL}}(N_\mathrm{b})]n_0+[ \mathcal{E}_{\text{X}}(N_\mathrm{t})-\mathcal{E}_{\text{X}}(N_\mathrm{t}-1))]\\
&-[ \mathcal{E}_{\text{X}}(N_\mathrm{b})-\mathcal{E}_{\text{X}}(N_\mathrm{b}-1))]\}/2E_0+2(N_\mathrm{t}-N_\mathrm{b}).
\end{split}
\end{equation}

Here the dimensionless $\widetilde{D}$ and the physical $D$ are related through Eq.~\eqref{criticalD}. $\widetilde{D}_{N_\mathrm{b},N_\mathrm{t}}^{(0)}$ is the displacement field at the center of the checkerboard labeled by $(N_\mathrm{b}, N_\mathrm{t})$. Equation~\eqref{criticalD} correctly captures the quantized interval of critical displacement fields within each Landau-level crossing checkerboard, as listed in Table~\ref{table:criticalfields}.

\begin{table}[h]
\caption{Critical displacement fields for $\widetilde{\nu}=1, 2, 3, 4$. The critical fields for $\widetilde{\nu}=5, 6, 7$ are the same as those for $\widetilde{\nu}=3, 2, 1$, respectively.  }
\centering
\begin{tabular}{c|c|c}
$\widetilde\nu=\widetilde\nu_{b}+\widetilde\nu_{t}  $&$ \left(\widetilde\nu_{b}, \widetilde\nu_{t}\right) \leftrightarrow\left(\widetilde\nu_{b}-1, \widetilde\nu_{t}+1\right)$  &$  \widetilde{D}_{N_{b}, N_{t}}^{(c)}-\widetilde{D}_{N_{b}, N_{t}}^{(0)}$  \\
\hline
$1$ & $(1,0) \leftrightarrow(0,1)  $&$ 0$\\
\hline 
\multirow{2}{*}{2} &$  (2,0) \leftrightarrow(1,1)  $&$  -\frac{1}{2} $\\
\cline{2-3}&$  (1,1) \leftrightarrow(0,2)  $&$  +\frac{1}{2} $\\ \hline
\multirow{3}{*}{3} &$  (3,0) \leftrightarrow(2,1)  $&$ -1$\\
\cline{2-3}&$  (2,1) \leftrightarrow(1,2)  $&$0$\\
\cline{2-3}&$  (1,2) \leftrightarrow(0,3)  $&$ +1$\\ 
\hline
\multirow{4}{*}{4} &$  (4,0) \leftrightarrow(3,1)  $&$  -\frac{3}{2} $\\
\cline{2-3}&$  (3,1) \leftrightarrow(2,2)  $&$  -\frac{1}{2} $\\
\cline{2-3}& $(2,2) \leftrightarrow(1,3)  $&$  +\frac{1}{2} $\\
\cline{2-3}&$  (1,3) \leftrightarrow(0,4)  $&$  +\frac{3}{2} $\\
\hline
\end{tabular}   
\label{table:criticalfields}
\end{table}

\textbf{III. Spin ordering and Zeeman effect} The effect of Zeeman energy, which is not taken into account in the above derivation, depends on the spin ordering at each filling factor. Meanwhile, the spin and valley ordering in the quantum Hall regime of graphene is determined by microscopic physics down to the atomic scale. Depending on the spin polarization at each filling factor, the Zeeman energy can shift the critical displacement fields and lead to corrections on the quantized interval. However, for Sample-S15 and S37, the intervals of critical displacement fields within each checkerboard display descent quantized value, as if the Zeeman energy does not play a role. One scenario for this observation is that the charge transfer between the two layers at the critical displacement fields is spin-conserved; the Zeeman energy of the system does not change across the transition, which leads to no correction on the critical displacement fields.

\section*{\label{sec:level1}Data Availability}

The data that support the findings of this study are available upon reasonable request to the corresponding authors.

\section*{\label{sec:level2}Code Availability}

The code that support the findings of this study are available upon reasonable request to the corresponding authors.

\section*{\label{sec:level3}Acknowledgements}
This work is supported by the National Key R$\&$D Program of China 2022YFA1203903, 2022YFA1402401, and the National Natural Science Foundation of China (NSFC) (Grant Nos. 92265203, 12034011, U23A6004, 12374185, U21A6004, 62204145 and 12274333). Z.V.H. acknowledges the support of the Fund for Shanxi “1331 Project” Key Subjects Construction. B.D and J.Z. acknowledge supports from the Innovation Program for Quantum Science and Technology (Grant No. 2021ZD0302003). K.W. and T.T. acknowledge support from the JSPS KAKENHI (Grant Numbers 20H00354 and 23H02052) and World Premier International Research Center Initiative (WPI), MEXT, Japan.

\section*{Author Contributions}
Z.H., and J.Z. (Jing Zhang) conceived the experiment and supervised the overall project.  B.D. and K.Z. performed the device fabrications and electrical measurements; F.W. carried out theoretical calculations; J.L. contributed to device fabrications; K.W. and T.T. provided high quality h-BN bulk crystals; Z.H., B.D., K.Z., F.W., J.L., J.Z. (Jianting Zhao) and J.Z. (Jing Zhang) analysed the experimental data. The manuscript was written by Z.H., B.D. and K.Z. with discussions and inputs from all authors.

\section*{Competing Interests}
The authors declare no competing interests.

\end{document}